\documentclass[preprintnumbers, prd, showpacs, floatfix,twocolumn,
preprintnumbers, letterpaper,superscriptaddress,nofootinbib]{revtex4}
\usepackage{amsfonts}
\usepackage{amsmath}
\usepackage{amssymb,epsf}
\usepackage{latexsym}
\usepackage{graphicx,epsfig}
\usepackage{amssymb}
\usepackage{subfigure}
\usepackage{epstopdf}
\usepackage[colorlinks=true,citecolor=blue,linkcolor=blue,urlcolor=black]{hyperref}

\begin{document}

\title{Thermodynamics of topological black holes in Brans-Dicke gravity \\
with a power-law Maxwell field}
\author{M. Kord Zangeneh}
\email{mkzangeneh@shirazu.ac.ir}
\affiliation{Physics Department and Biruni Observatory, College of Sciences, Shiraz
University, Shiraz 71454, Iran}
\author{M. H. Dehghani}
\email{mhd@shirazu.ac.ir}
\author{A. Sheykhi}
\email{asheykhi@shirazu.ac.ir}
\affiliation{Physics Department and Biruni Observatory, College of Sciences, Shiraz
University, Shiraz 71454, Iran}
\affiliation{Research Institute for Astronomy and Astrophysics of Maragha (RIAAM), P.O.
Box 55134-441, Maragha, Iran}

\begin{abstract}
In this paper, we present a new class of higher-dimensional exact
topological black hole solutions of the Brans-Dicke theory in the presence
of a power-law Maxwell field as the matter source. For this aim, we
introduce a conformal transformation which transforms the
Einstein-dilaton-power-law Maxwell gravity Lagrangian to the
Brans-Dicke-power-law Maxwell theory one. Then, by using this conformal
transformation, we obtain the desired solutions. Next, we study the
properties of the solutions and conditions under which we have black holes.
Interestingly enough, we show that there is a cosmological horizon in the
presence of a negative cosmological constant. Finally, we calculate the
temperature and charge and then by calculating the Euclidean action, we
obtain the mass, the entropy and the electromagnetic potential energy. We
find that the entropy does not respect the area law, and also the conserved
and thermodynamic quantities are invariant under conformal transformation.
Using these thermodynamic and conserved quantities, we show that the first
law of black hole thermodynamics is satisfied on the horizon.
\end{abstract}

\maketitle

\section{INTRODUCTION\label{Intro}}

During the years between $1905$ and $1915$, many attempts were done to enter
gravity into special relativity. Einstein, himself, and Nordstrom were
pioneers in this issue. Nordstrom presented a scalar theory of gravitation 
\cite{Nor} while Einstein proposed a tensorial theory of gravitation \cite%
{Ein}. Eventually, the Einstein tensorial theory of gravitation, known today
as general relativity (GR), was more successful to pass observational tests
and became the standard theory of gravitation \cite{Laue}. However, some
time later, GR showed some failures. The most important failure was the
inability to describe accelerating expansion of the Universe \cite{Expan}.
In addition to latter failure, the fact that Mach's principle is not
respected by GR led physicists to modify Einstein theory of gravity. One of
the most impressive and physically viable modifications is Brans-Dicke (BD)
theory adding a scalar degree of freedom to previous 10 degrees of freedom
coming from metric tensor $g_{\mu \nu }$ in four dimensions \cite{BD}. This
theory was motivated from one side by Mach's principle encoded by varying
gravitational constant in it and from other side by Dirac's large number
hypothesis. These two motivations are summed up in this theory by
considering a scalar field, $\Phi $, which is inversely proportional to the
gravitational constant $G$ and nonminimally coupled to gravitation \cite%
{Brans1}. Another important aspect of BD theory is its equivalence to
several other modifications of GR in particular $f(R)$ theories that causes
better understating of these theories \cite{f(R)}. Also, BD theory may be
considered as a special case of more general scalar-tensor gravity which has
been considered in \cite{nonlin}.

The first solutions of BD theory have been published by Brans himself. These
solutions were in four dimensions and classified in four classes \cite%
{Brans2}. Among these four classes just two of them are really independent 
\cite{Bhadra} while just one of these two independent classes are valid for
all values of $\omega $. Also, linearly charged solutions of BD gravity have
been presented in \cite{Cai}. These latter solutions are just allowed in the
presence of a trivial (constant) scalar field in four dimensions. This is
due to the conformal invariance of linear Maxwell Lagrangian in four
dimensions. Since linear Maxwell Lagrangian is no longer conformally
invariant in higher dimensions, the linear Maxwell field can play the role
of source of scalar field for higher-dimensional BD gravity \cite{Cai}. The
highly nonlinear nature of BD theory makes it non-straightforward in many
cases to find the solutions by solving BD field equations directly.
Fortunately, there is a way to overcome this problem. As it has been shown
in many cases, the solutions of BD theory can be found by applying a
conformal transformation on known solutions of other theories such as
dilaton theory \cite{Confsol1}. For instance, linearly charged rotating
black branes have been obtained in BD theory by applying a conformal
transformation from the known solutions of dilaton gravity \cite{Confsol4}.
Also, by using a similar method, asymptotic anti-de Sitter (AdS) black holes
and topological black holes with nonflat and non-AdS asymptotic behaviors
have been investigated in Refs. \cite{Confsol2} and \cite{Confsol3},
respectively. The fact that one can transform the Lagrangian of BD gravity
to the Lagrangian of dilaton gravity by a conformal transformation has been
also used for more general scalar-tensor gravities in \cite{nonlin}to
investigate the different aspects of scalar-tensor gravities in the presence
of nonlinear electromagnetic field \cite{nonlin}.

The nonlinear electrodynamics was first introduced by Born and Infeld in
order to obtain the finite energy density for an electron \cite{BI}. In
recent years other types of Born-Infeld like nonlinear electrodynamics have
been proposed \cite{Soleng, HendiJHEP}. The studies were also extended to
dilaton gravity \cite{Shey}. However, the energy-momentum tensors of
Born-Infeld theory or the Born-Infeld like theories are not traceless even
in four dimensions. A nonlinear electrodynamic theory with a traceless
energy-momentum tensor in higher dimensions has been introduced in \cite%
{PLM1}. The power-law Maxwell matter source is conformal invariant in higher
dimensions for special choice of power. To be more clear, the Lagrangian of
power-law Maxwell field $\left( -F_{\mu \nu }F^{\mu \nu }\right) ^{p}$ is
invariant under conformal transformation $g_{\mu \nu }\rightarrow \Omega
^{2}g_{\mu \nu }$, $A_{\mu }\rightarrow A_{\mu }$ provided $p=(n+1)/4$.
Despite of the mentioned special property for $p=(n+1)/4$, many solutions
have been studied from different aspects with nonfixed $p$ \cite{NFP}. In
this paper we extend the study of \cite{Confsol3} to nonlinear power-law
Maxwell field electrodynamics and try to solve field equations of BD theory
by using known solutions of dilaton gravity that we have recently presented
in \cite{KSD}.

The layout of the paper is as follows. In the next section, we introduce the
basic field equations of BD theory with power-law Maxwell Lagrangian. We
also introduce a conformal transformation which transforms this theory to
Einstein-dilaton gravity. In Sec. \ref{Sol}, we obtain a class of
topological black hole solution of this theory and investigate its
properties. In Sec. \ref{Ther}, we investigate thermodynamics of the
solutions and check the validity of the first law of thermodynamics. We
finish our paper with closing remarks in the last section.

\section{FIELD EQUATIONS AND CONFORMAL TRANSFORMATION \label{Con}}

The action of Brans-Dicke theory coupled to a power-law Maxwell (BDPM) field
can be written as%
\begin{eqnarray}
S_{BD} &=&-\frac{1}{16\pi }\int_{\mathcal{M}}d^{n+1}x\sqrt{-g}\left( \Phi 
\mathcal{R}\text{ }-\frac{\omega }{\Phi }(\nabla \Phi )^{2}\right.  \notag \\
&&\text{ \ \ \ \ \ \ \ \ \ \ \ \ \ \ \ \ \ \ \ \ \ \ \ \ \ \ \ \ }\left.
-V(\Phi )+(-F)^{p}\right) ,  \label{Act}
\end{eqnarray}%
where $\mathcal{R}$ is the Ricci scalar, $\omega $ is the coupling constant, 
$\Phi $ denotes the BD scalar field, $p$ is a constant determining the
nonlinearity of the electromagnetic field and $V(\Phi )$ is a
self-interacting potential for $\Phi $. In Eq. (\ref{Act}), $F=F_{\mu \nu
}F^{\mu \nu }$ where $F_{\mu \nu }=\partial _{\lbrack \mu }A_{\nu ]}$ is the
electromagnetic tensor field and $A_{\mu }$ is the vector potential. Varying
the action (\ref{Act}) with respect to the metric $g_{\mu \nu }$, the scalar
field $\Phi $ and the electromagnetic field $A_{\mu }$, one can obtain the
field equations as 
\begin{eqnarray}
G_{\mu \nu } &=&\frac{\omega }{\Phi ^{2}}\left( \nabla _{\mu }\Phi \nabla
_{\nu }\Phi -\frac{1}{2}g_{\mu \nu }(\nabla \Phi )^{2}\right)  \notag \\
&&+\frac{1}{\Phi }\left( \nabla _{\mu }\nabla _{\nu }\Phi -g_{\mu \nu
}\nabla ^{2}\Phi \right)  \notag \\
&&-\frac{V(\Phi )}{2\Phi }g_{\mu \nu }-\frac{2(-F)^{p-1}}{\Phi }\left( \frac{%
1}{4}Fg_{\mu \nu }-pF_{\mu \lambda }F_{\nu }^{\text{ }\lambda }\right) , 
\notag \\
&&  \label{fil1} \\
\nabla ^{2}\Phi &=&\frac{(n-4p+1)(-F)^{p}}{2\left[ \left( n-1\right) \omega
+n\right] }  \notag \\
&&+\frac{\left[ (n-1)\Phi \frac{dV(\Phi )}{d\Phi }-\left( n+1\right) V(\Phi )%
\right] }{2\left[ \left( n-1\right) \omega +n)\right] },  \label{fil2} \\
&&\nabla _{\mu }\left[ (-F)^{p-1}F^{\mu \nu }\right] =0,  \label{fil3}
\end{eqnarray}%
As it is clear from the right-hand side of (\ref{fil1}), there are second
order derivatives of the scalar field. This fact makes the problem of
solving the field equations (\ref{fil1})-(\ref{fil3}) difficult.
Fortunately, this difficulty can be circumvented by using the following
conformal transformation: 
\begin{gather}
\tilde{\Phi}=\frac{\sqrt{n+\omega (n-1)}}{2}\ln \Phi ,  \notag \\
\tilde{g}_{\mu \nu }=\Phi ^{2/(n-1)}g_{\mu \nu },  \notag \\
F_{\mu \nu }=\tilde{F}_{\mu \nu },  \label{Conf}
\end{gather}%
where $\tilde{\Phi}$ is the dilaton field and 
\begin{equation}
\tilde{V}(\tilde{\Phi})=\Phi ^{-(n+1)/(n-1)}V(\Phi ).
\end{equation}%
Indeed, the transformation (\ref{Conf}) transforms the action (\ref{Act}) to
the action of Einstein-dilaton gravity coupled to power-law Maxwell field 
\cite{KSD} 
\begin{eqnarray}
S_{ED} &=&-\frac{1}{16\pi }\int d^{n+1}x\sqrt{-\tilde{g}}\left\{ \mathcal{%
\tilde{R}}-\frac{4}{n-1}(\tilde{\nabla}\tilde{\Phi})^{2}\right.  \notag \\
&&\left. -\tilde{V}(\tilde{\Phi})+\left( -e^{-4\alpha \tilde{\Phi}/(n-1)}%
\tilde{F}\right) ^{p}\right\} ,  \label{Act2}
\end{eqnarray}%
where $\alpha =(n-4p+1)/2p\sqrt{n+\omega (n-1)}$. One may note that $\alpha $
is a constant which determines the strength of coupling of the scalar and
electromagnetic field and is equal to zero for $p=(n+1)/4$. So, we assume
that $p\neq (n-1)/4$. The field equations corresponding to action (\ref{Act2}%
) are 
\begin{eqnarray}
\mathcal{\tilde{R}}_{\mu \nu } &=&\tilde{g}_{\mu \nu }\left\{ \frac{\tilde{V}%
(\tilde{\Phi})}{n-1}+\frac{(2p-1)}{n-1}\left( -e^{-4\alpha \tilde{\Phi}%
/(n-1)}\tilde{F}\right) ^{p}\right\}  \notag \\
&&+\frac{4}{n-1}\partial _{\mu }\tilde{\Phi}\partial _{\nu }\tilde{\Phi}%
+2pe^{-4\alpha p\tilde{\Phi}/(n-1)}(-\tilde{F})^{p-1}\tilde{F}_{\mu \lambda }%
\tilde{F}_{\nu }^{\text{ \ }\lambda },  \notag \\
&&  \label{FE1}
\end{eqnarray}%
\begin{eqnarray}
\tilde{\nabla}^{2}\tilde{\Phi}-\frac{n-1}{8}\frac{d\tilde{V}(\tilde{\Phi})}{d%
\tilde{\Phi}}-\frac{p\alpha }{2}e^{-{4\alpha p\tilde{\Phi}}/({n-1})}(-\tilde{%
F})^{p} &=&0,  \label{FE2} \\
\tilde{\nabla}_{\mu }\left( e^{-{4\alpha p\tilde{\Phi}}/({n-1})}(-\tilde{F}%
)^{p-1}\tilde{F}^{\mu \nu }\right) &=&0.  \label{FE3}
\end{eqnarray}%
One should note that the field equation (\ref{FE1}) does not contain second
order derivatives of scalar field and, therefore, one can solve Eqs. (\ref%
{FE1})-(\ref{FE2}) in a simpler way than the field equations (\ref{fil1})-(%
\ref{fil3}). In \cite{KSD}, the solution of Einstein-dilaton-power Maxwell
(EDPM) gravity has been presented. We first review these solutions in the
next section.

\section{TOPOLOGICAL BLACK HOLE SOLUTIONS IN BDPM GRAVITY \label{Sol}}

In order to obtain the topological black hole solutions of BDPM gravity, we
first review the $(n+1)$-dimensional topological black hole solutions of
EDPM gravity (\ref{Act2}) presented in \cite{KSD}. The metric of a general
static spacetime can be written as 
\begin{equation}
d\tilde{s}^{2}=-f(r)dt^{2}+{\frac{dr^{2}}{f(r)}}+r^{2}R^{2}(r)d{\Omega }%
_{n-1}^{2},  \label{metric}
\end{equation}%
where $d\Omega _{n-1}^{2}=h_{ij}(x)dx^{i}dx^{j}$ is the metric of an $(n-1)$%
-dimensional constant curvature hypersurface. The curvature of this
hypersurface is equal to $(n-1)(n-2)k$, where $k=0,\pm 1$. We have shown in 
\cite{KSD} that in order to have topological black holes with a general $k$
and $p$, the potential should be chosen as

\begin{equation}
\tilde{V}(\tilde{\Phi})=2\,\Lambda _{1}e{^{2\zeta _{1}\tilde{\Phi}}}%
+2\Lambda _{2}e{^{2\zeta _{2}\tilde{\Phi}}}+2\,\Lambda e{^{2\zeta _{3}\tilde{%
\Phi}}},  \label{v2}
\end{equation}%
where%
\begin{gather}
\zeta _{1}={\frac{2}{\left( n-1\right) \alpha }},\hspace{0.8cm}\zeta _{2}={%
\frac{2p\left( n-1+{\alpha }^{2}\right) }{\left( n-1\right) \left(
2\,p-1\right) \alpha }},  \notag \\
\zeta _{3}=\,{\frac{2\alpha }{n-1}},\hspace{0.8cm}\Lambda _{1}={\frac{%
k\left( n-1\right) \left( n-2\right) {\alpha }^{2}}{2{b}^{2}\left( {\alpha }%
^{2}-1\right) }},  \notag \\
\Lambda _{2}=\frac{2^{p-1}\left( 2\,p-1\right) \left( p-1\right) {\alpha }%
^{2}\,{q}^{2\,p}}{\Pi {b}^{{\frac{2\left( n-1\right) p}{2\,p-1}}}}.
\end{gather}
The solutions of EDPM field equations are \cite{KSD}

\begin{eqnarray}
f(r) &=&\frac{k\left( n-2\right) (1+\alpha ^{2})^{2}{r}^{2\gamma }}{%
(1-\alpha ^{2})\left( {\alpha }^{2}+n-2\right) {b}^{2\gamma }}-\frac{m}{{r}%
^{(n-1)(1-\gamma )-1}}  \notag \\
&&+\frac{\hat{q}^{2p}\,}{{r}^{\Upsilon +\left( n-1\right) (1-\gamma )-1}}-%
\frac{2\Lambda {b}^{2\gamma }(1+\alpha ^{2})^{2}{r}^{2(1-\gamma )}}{\left(
n-1\right) \left( n-{\alpha }^{2}\right) },  \notag \\
&&  \label{f(r)}
\end{eqnarray}

\begin{equation}
\tilde{\Phi}(r)=\frac{\,\left( n-1\right) \alpha }{2\left( {\alpha }%
^{2}+1\right) }\,\ln \left( {\frac{b}{r}}\right) ,  \label{phi}
\end{equation}%
\begin{equation}
R(r)=\left( {\frac{b}{r}}\right) ^{\gamma },  \label{Rphi}
\end{equation}%
\begin{equation}
\tilde{A}_{t}=\frac{q{b}^{{\frac{\left( 2\,p+1-n\right) \gamma }{\left(
2\,p-1\right) }}}}{\Upsilon {r}^{\Upsilon }},  \label{At}
\end{equation}%
where 
\begin{eqnarray*}
\hat{q}^{2p} &=&\frac{2^{p}p\left( 2p-1\right) {q}^{2\,p}\,}{(1-\gamma )\Pi
\Upsilon {b}^{{2\left( n-2\right) p\gamma /}\left( 2\,p-1\right) }}, \\
\Pi &=&{\alpha }^{2}+\left( n-1-{\alpha }^{2}\right) p, \\
\Upsilon &=&\frac{{(n-2p+\alpha }^{2}{)}}{{(2p-1)(1+\alpha }^{2})},
\end{eqnarray*}%
$b$ is an arbitrary nonzero positive constant, and $\gamma =\alpha
^{2}/(\alpha ^{2}+1)$. In the above relations $m$ and $q$ are two constants
proportional to the mass and charge of the black holes, respectively, and $%
\Lambda $ is a free constant which can be interpreted as cosmological
constant, since in the absence of the dilaton field $(\alpha =0)$, we have $%
V(\Phi )=2\Lambda $. Thus as usual, we redefine it as $\Lambda
=-n(n-1)/2l^{2}$ where $l$ is a constant with length dimension.

With the solutions of EDPM gravity in hand, we can construct $(n+1)$%
-dimensional solutions of BDPM by applying conformal transformation (\ref%
{Conf}). The line element of the spacetime can be obtained as%
\begin{equation}
ds^{2}=-U(r)dt^{2}+{\frac{dr^{2}}{V(r)}}+r^{2}H^{2}(r)h_{ij}dx^{i}dx^{j},
\label{metric2}
\end{equation}%
where $U(r)$, $V(r)$, $H(r)$ and $\Phi (r)$ are

\begin{eqnarray}
U(r) &=&\left( {\frac{b}{r}}\right) ^{-\Gamma }f(r)  \notag \\
&=&\frac{k\left( n-2\right) (1+\alpha ^{2})^{2}}{(1-\alpha ^{2})\left( {%
\alpha }^{2}+n-2\right) }\left( {\frac{r}{b}}\right) ^{\Gamma +2\gamma } 
\notag \\
&&-\frac{m{b}^{-\Gamma }}{{r}^{-\Gamma +(n-1)(1-\gamma )-1}}+\frac{%
b^{-\Gamma }\hat{q}^{2p}}{r^{-{\Gamma }+\Upsilon +\left( n-1\right)
(1-\gamma )-1}}  \notag \\
&&-\frac{2\Lambda b^{2}(1+\alpha ^{2})^{2}}{\left( n-1\right) \left( n-{%
\alpha }^{2}\right) }\left( {\frac{r}{b}}\right) ^{\Gamma +2(1-\gamma )},
\label{U(r)}
\end{eqnarray}

\begin{eqnarray}
V(r) &=&\left( {\frac{b}{r}}\right) ^{\Gamma }f(r)  \notag \\
&=&\frac{k\left( n-2\right) (1+\alpha ^{2})^{2}}{(1-\alpha ^{2})\left( {%
\alpha }^{2}+n-2\right) }\left( {\frac{r}{b}}\right) ^{-\Gamma +2\gamma } 
\notag \\
&&-\frac{m{b}^{\Gamma }}{{r}^{\Gamma +(n-1)(1-\gamma )-1}}+\frac{{b}^{{%
\Gamma }}\hat{q}^{2p}}{r^{{\Gamma }+\Upsilon +\left( n-1\right) (1-\gamma
)-1}}  \notag \\
&&-\frac{2\Lambda {b}^{2}(1+\alpha ^{2})^{2}}{\left( n-1\right) \left( n-{%
\alpha }^{2}\right) }\left( {\frac{r}{b}}\right) ^{-\Gamma +2(1-\gamma )},
\label{V(r)}
\end{eqnarray}

\begin{eqnarray}
H(r) &=&\left( {\frac{b}{r}}\right) ^{\gamma -\Gamma /2},  \label{H(r)} \\
\Phi (r) &=&\left( {\frac{b}{r}}\right) ^{\left( n-1\right) \Gamma /2},
\label{Phi2}
\end{eqnarray}%
where $\Gamma =4p\gamma /\left( n-4p+1\right) $. The electromagnetic gauge
potential $A_{t}$ and potential of scalar field can also be obtained as

\begin{equation}
A_{t}=\frac{q{b}^{{\frac{\left( 2\,p+1-n\right) \gamma }{\left(
2\,p-1\right) }}}}{\Upsilon {r}^{\Upsilon }},
\end{equation}

\begin{equation}
V(\Phi )=2\,\Lambda _{1}\Phi ^{\xi _{1}}+2\Lambda _{2}\Phi ^{\xi
_{2}}+2\,\Lambda \Phi ^{\xi _{3}},  \label{V}
\end{equation}%
where%
\begin{gather*}
\xi _{1}={\frac{n-4p+1+p{\alpha }^{2}(n+1)}{p\left( n-1\right) {\alpha }^{2}}%
,}\text{ \ \ \ \ }\xi _{2}=\frac{n-4p+1+2p\alpha ^{2}}{\alpha ^{2}\left(
2p-1\right) }, \\
\xi _{3}=\frac{p(n-3)+n+1}{p\left( n-1\right) }.
\end{gather*}%
It is worth mentioning that for the case of $p=1$ and $k=0$, both $\Lambda
_{1}$ and $\Lambda _{2}$ vanish and therefore $V(\Phi )=2$, $\Lambda \Phi
^{2}$ \cite{Confsol4}. Also, one may note that as $\omega \rightarrow \infty 
$ ($\alpha =\gamma =0$) for the linear Maxwell theory ($p=1$), solutions (%
\ref{U(r)}) and (\ref{V(r)}) reduce to%
\begin{equation}
U(r)=V(r)=k-\frac{m}{r^{n-2}}+\frac{2q^{2}}{(n-1)(n-2)r^{2(n-2)}}-\frac{%
2\Lambda }{n(n-1)}r^{2},
\end{equation}%
which describes an $(n+1)$-dimensional asymptotically (A)dS topological
black hole for positive (negative) cosmological constant with a flat ($k=0$%
), spherical ($k=1$) or hyperbolic ($k=-1$) horizons (see for example \cite%
{Bril1,Cai3,lemos}). For $p=1$, our solutions reproduce the solutions of
Ref. \cite{Confsol3}.

\begin{figure}[t]
\epsfxsize=7cm \centerline{\epsffile{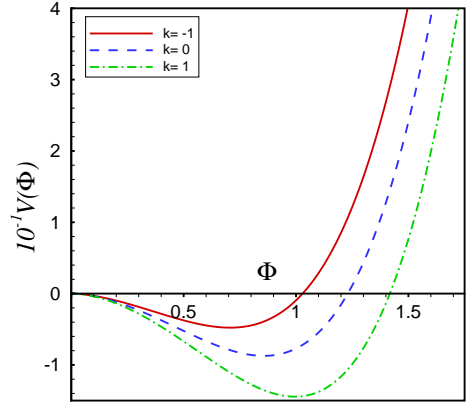}}
\caption{The self-interacting potential $10^{-1}V(\Phi )$ versus $\Phi $ for 
$n=5$, $p=1.2$, $\protect\alpha =0.6$, $l=b=1$, $\Lambda =-10$ and $q=10$.}
\label{fig1}
\end{figure}

One should note that there are some constraints on the values of $p$ and $%
\alpha $. We stress on this fact that the electromagnetic gauge potential $%
A_{t}$ should vanish at infinity and, therefore, $\Upsilon >0$ which leads to

\begin{equation}
\frac{1}{2}<p<\frac{n+\alpha ^{2}}{2}.  \label{res1}
\end{equation}%
On the other hand, we assume that the scalar field of BD be a localized
function. That is $\Phi $ should go to zero as $r\rightarrow \infty $ which
implies $\Gamma >0$ and therefore

\begin{equation}
p<\frac{n+1}{4}.  \label{res2}
\end{equation}
One can summarize (\ref{res1}) and (\ref{res2}) to receive

\begin{equation}
\frac{1}{2}<p<\frac{n+1}{4},  \label{res3}
\end{equation}%
which is the allowed range of $p$. In the above allowed range of $p$ the
charge term of the metric functions $U(r)$ given in Eq. (\ref{U(r)})
dominates in the vicinity of $r=0$. Consequently, it is sufficient to check
the sign of the power of $r$ in the mass term in $U(r)$ in order to
guarantee that the effects of mass and charge vanish at infinity in both $%
U(r)$ and $V(r)$ given by (\ref{U(r)}) and (\ref{V(r)}) [note that according
to (\ref{res3}), $\Gamma >0$]. One can easily show that the effects of mass
in $U(r)$ disappear at infinity provided

\begin{equation}
{\alpha }^{2}<\frac{(n-4p+1)(n-2)}{n+1}.  \label{res4}
\end{equation}%
or equivalently 
\begin{equation}
\omega >-\frac{n}{n-1}+\frac{(n-4p+1)(n+1)}{4p^{2}(n-1)(n-2)}.
\end{equation}%
Hence the allowed ranges of $p$ and $\alpha $ or $\omega $ are 
\begin{equation}
\frac{1}{2}<p<\frac{n+1}{4},\text{ \ \ \ \ }{\alpha }^{2}<\frac{(n-4p+1)(n-2)%
}{n+1}  \label{res5}
\end{equation}%
or equivalently%
\begin{equation}
\frac{1}{2}<p<\frac{n+1}{4},\text{ \ \ \ \ }\omega >-\frac{n}{n-1}+\frac{%
(n-4p+1)(n+1)}{4p^{2}(n-1)(n-2)}.
\end{equation}
Before discussing the properties of our solutions, we pause to consider the
stability of the system. In order to have a stable system, it is necessary
for $V(\Phi )$ to have a lower bound. As it can be seen from Fig. \ref{fig1}%
, $V(\Phi )$ is bounded from below for suitable choices of the parameters. 

\subsection{Properties of the solutions}

Now, we are ready to discuss physical properties of the topological BD black
holes in the presence of power-law Maxwell field. First, we discuss the
domain of validity of our solutions. It is worth mentioning that, as one can
see from Eqs. (\ref{U(r)}) and (\ref{V(r)}), our solutions are ill defined
for the string case where $\alpha =1$ [corresponding to the case of $\omega
=[(n-4p+1)^{2}-4np^{2}]/4p^{2}(n-1)$], except for the flat horizon case i.e. 
$k=0$. For other values of $\alpha $, our solutions are well defined in the
permitted ranges of $p$ and $\alpha $. Second, it is easy to show that there
is an essential singularity at $r=0$ since the Kretschmann scalar $R_{\mu
\nu \lambda \sigma }R^{\mu \nu \lambda \sigma }$ diverges at $r=0$.
Kretschmann scalar also is finite for $r\neq 0$ and vanishes as $%
r\rightarrow \infty .$ Third, the charge term in both $V(r)$ and $U(r)$ is
the dominant term in the vicinity of $r=0$ and goes to infinity for the
allowed ranges of $p$ and $\alpha $ ($\hat{q}^{2p}>0$). Thus, the
singularity is timelike as in the case of Reissner-Nordstrom black holes.
That is, there is no Schwarzschild-like solution and one encounters with the
solutions with two inner and outer horizons, extreme black holes and naked
singularities depending on the values of the metric parameters such as $p$, $%
q$, $m$, $\alpha $ and $k$. Fourth, we investigate the asymptotic behavior
of the solutions. In order to consider the asymptotic behavior of our
solutions, it is enough to investigate the behavior of the metric function $%
V(r)$ at infinity. In the case of $k=0$, for $\alpha ^{2}<\min
[(n-4p+1)/2p,(n-4p+1)(n-2)/(n+1)]$ [Note that $\alpha ^{2}<(n-4p+1)/2p$
implies that $-\Gamma +2(1-\gamma )>0$ and ${\alpha }%
^{2}<(n-4p+1)(n-2)/(n+1) $ is (\ref{res4})], the metric functions go to
infinity as $r$ goes to infinity provided $\Lambda <0$, and one has
cosmological horizon for $\Lambda >0$. In the case of $k=\pm 1$, for $\alpha
^{2}>1$, the first term of $V(r)$ is the dominant term at infinity.
Therefore, the metric functions go to infinity in the case of $k=-1$ and one
has cosmological horizon in the case of $k=1$ provided $p<(n+1)/6$ ($-\Gamma
+2\gamma >0$). For $\alpha ^{2}<1$ where the fourth term of $V(r)$ is
dominant at infinity, the metric functions go to infinity for $\Lambda <0$\
and the solutions have cosmological horizon for $\Lambda >0$ provided $%
\alpha ^{2}<\min \left[ (n-4p+1)/2p,(n-4p+1)(n-2)/(n+1)\right] $.

Finally, we investigate the causal structure of our solutions. In order to
do this, and find out whether the singularity is naked or not, we should
study the zeros of $g^{rr}=V(r)$. Although it is difficult to find the roots
of $V(r)=0$ analytically, we can gain some insight into the behavior of it
by studying $m(r_{h})$:%
\begin{eqnarray}
m(r_{h}) &=&\frac{k\left( n-2\right) (1+\alpha ^{2})^{2}{b}^{-2\gamma }}{%
(1-\alpha ^{2})\left( {\alpha }^{2}+n-2\right) }{r}_{h}^{(n-3)(1-\gamma )+1}
\notag \\
&&+\frac{\hat{q}^{2p}}{r_{h}^{\Upsilon }}-\frac{2\Lambda {b}^{2\gamma
}(1+\alpha ^{2})^{2}}{\left( n-1\right) \left( n-{\alpha }^{2}\right) }{r}%
_{h}^{(n+1)(1-\gamma )-1},  \notag \\
&&
\end{eqnarray}%
\begin{figure}[t]
\epsfxsize=7cm \centerline{\epsffile{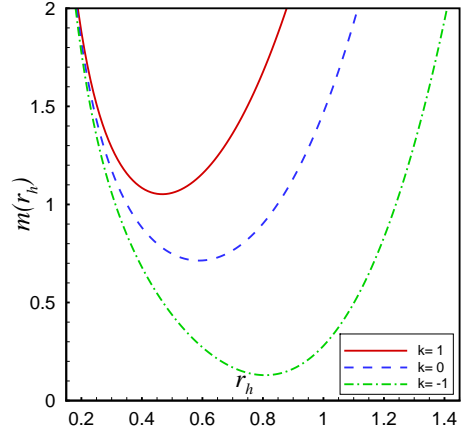}}
\caption{The function $m(r_{h})$ versus $r_{h}$ for $n=4$, $p=1.2$, $\protect%
\alpha =0.26$, $l=b=1$, $\Lambda =-6$ and $q=0.6$.}
\label{fig2}
\end{figure}
\begin{figure}[t]
\epsfxsize=7cm \centerline{\epsffile{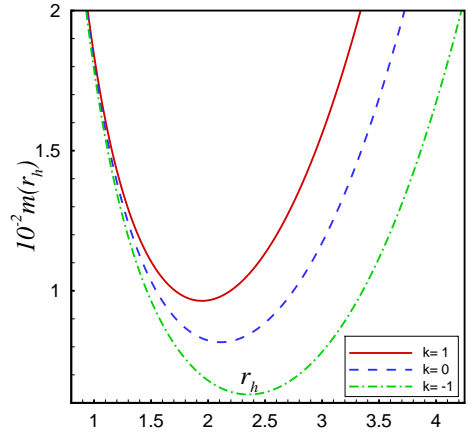}}
\caption{The function $10^{-2}m(r_{h})$ versus $r_{h}$ for $n=5$, $p=1.2$, $%
\protect\alpha =0.6$, $l=b=1$, $\Lambda =-10$ and $q=10$.}
\label{fig3}
\end{figure}
which comes from this fact that $V(r_{h})=0$ where $r_{h}$ is the radius of
the horizon. We consider $m(r_{h})$ in the absence of the cosmological
horizon and with cosmological horizon, separately. In the absence of
cosmological horizon, as $r_{h}$ goes from zero to infinity, the function $%
m(r_{h})$ starts from infinity and goes to infinity as one can see in Figs. %
\ref{fig2} and \ref{fig3}. The intersections of the line $m=\mathrm{constant}
$ with the curve $m(r_{h})$ are inner and outer horizons $r_{-}$ and $r_{+}$
in these figures. There is also a minimum in these figures. The value of the
minimum is $m_{\mathrm{ext}}$ which is the solution of $V(r)=0=V^{\prime
}(r) $: 
\begin{eqnarray}
m_{\mathrm{ext}} &=&\frac{2\left( {\alpha }^{2}+1\right) }{{(2p-1)}\Upsilon }%
{r}_{\mathrm{ext}}^{({n+1)(1-\gamma )-1}}  \notag \\
&&\times \left[ -\frac{2\Lambda {b}^{2\gamma }\Pi }{\left( n-{\alpha }%
^{2}\right) \left( n-1\right) }\right.  \notag \\
&&\left. +\frac{k\left( \left( n-3+{\alpha }^{2}\right) p+1\right) \left(
n-2\right) }{{b}^{2{\gamma }}\left( 1-\alpha ^{2}\right) \left( {\alpha }%
^{2}+n-2\right) }{r}_{\mathrm{ext}}^{4\gamma -2}\right] .  \label{mext}
\end{eqnarray}%
\begin{figure}[t]
\epsfxsize=7cm \centerline{\epsffile{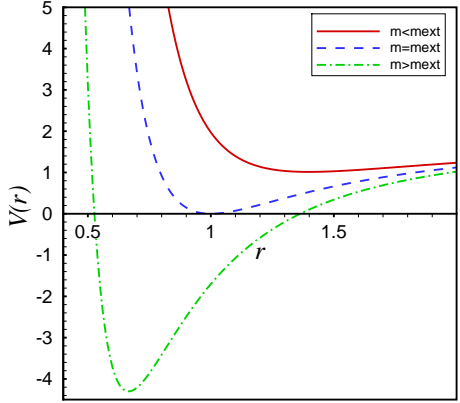}}
\caption{The function $V(r)$ versus $r$ for $k=0$, $n=4$, $p=1.2$, $\protect%
\alpha =0.26$, $l=b=1$, $\Lambda =-6$ and $r_{\mathrm{ext}}=1$.}
\label{fig4}
\end{figure}
\begin{figure}[t]
\epsfxsize=7cm \centerline{\epsffile{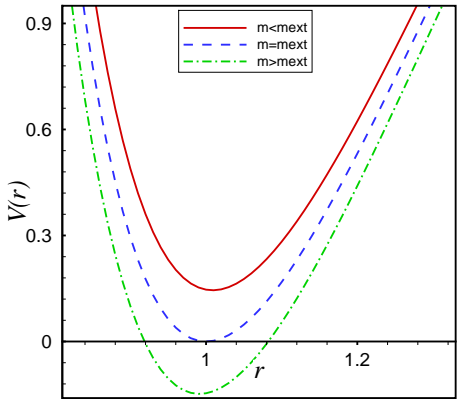}}
\caption{The function $V(r)$ versus $r$ for $k=-1$, $n=6$, $p=1.1$, $\protect%
\alpha =1.18$, $l=b=1$, $\Lambda =15$ and $r_{\mathrm{ext}}=1$.}
\label{fig5}
\end{figure}
\begin{figure}[t]
\epsfxsize=7cm \centerline{\epsffile{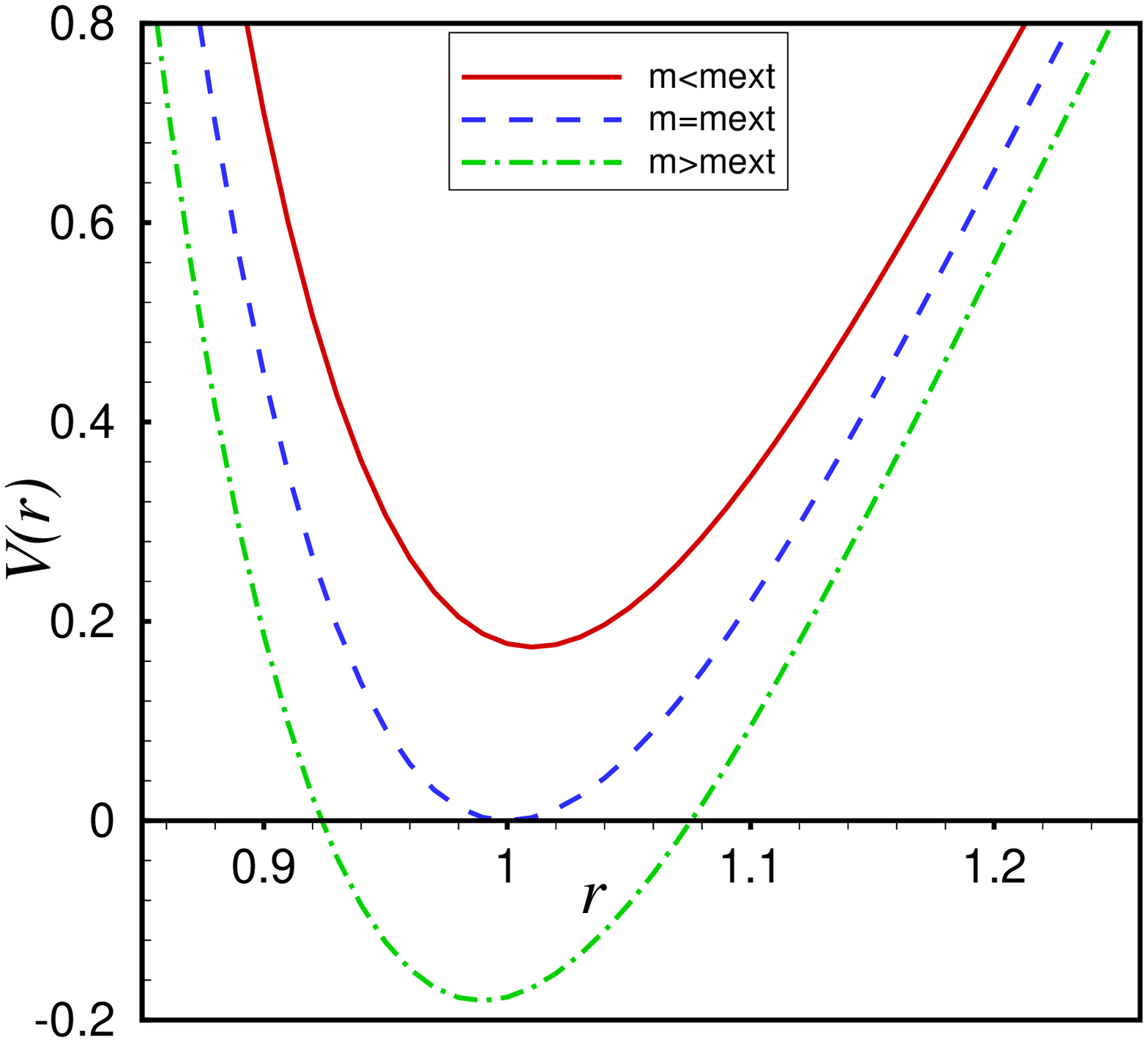}}
\caption{The function $V(r)$ versus $r$ for $k=1$, $n=5$, $p=1.2$, $\protect%
\alpha =0.63$, $l=b=1$, $\Lambda =-10$ and $r_{\mathrm{ext}}=1$.}
\label{fig6}
\end{figure}
\begin{figure}[t]
\epsfxsize=7cm \centerline{\epsffile{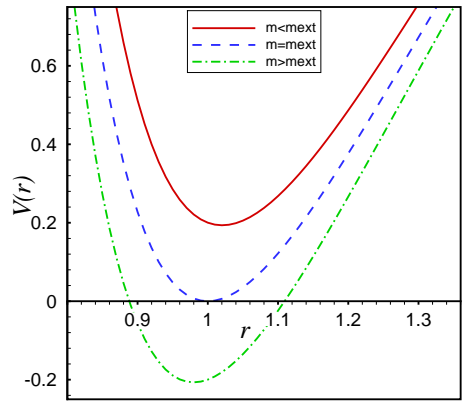}}
\caption{The function $V(r)$ versus $r$ for $k=-1$, $n=5$, $p=1.2$, $\protect%
\alpha =0.63$, $l=1$, $b=1.5$, $\Lambda =-10$ and $r_{\mathrm{ext}}=1$.}
\label{fig7}
\end{figure}
Thus, our solutions present black holes with inner and outer horizons
located at $r_{-}$ and $r_{+}$ provided $m>m_{\mathrm{ext}}$, an extreme
black hole if $m=m_{\mathrm{ext}}$ and a naked singularity provided $m<m_{%
\mathrm{ext}}$ (See Figs. \ref{fig4}-\ref{fig7}). For solutions with
cosmological horizon, the function $m(r_{h})$ starts from infinity and goes
to minus infinity as $r_{h}$ increases from zero to infinity. In this case
it is better to study the behavior of $\partial m/\partial r_{h}$ which is
given as 
\begin{eqnarray}
\frac{\partial m}{\partial r_{h}} &=&\frac{k\left( {\alpha }^{2}+1\right)
\left( n-2\right) }{\left( 1-{\alpha }^{2}\right) {b}^{2\gamma }}{r}_{h}^{{%
(n-3)(1-\gamma )}}  \notag \\
&&-\frac{\Upsilon \hat{q}^{2p}}{{r}_{h}^{{\Upsilon +1}}}-\frac{2\Lambda
\left( {\alpha }^{2}+1\right) {b}^{2\gamma }}{\left( n-1\right) }{r}%
_{h}^{(n+1)(1-\gamma )-2}.  \label{dmdr}
\end{eqnarray}%
As it is clear from Eq. (\ref{dmdr}), $\partial m/\partial r_{h}$ has no
zero for $k/\left( 1-{\alpha }^{2}\right) \leq 0$ and $\Lambda >0$ and the
function $m(r_{h})$ starts at infinity and goes to minus infinity without
any minimum or maximum. For this case our solution presents a naked
singularity with cosmological horizon as Fig. \ref{fig8} shows. When $%
\partial m/\partial r_{h}$ has zeros, the function $m(r_{h})$ starts at
infinity and goes to minus infinity with a minimum and a maximum as one can
see in Fig. \ref{fig9}. In this case we have black holes with three inner,
outer and cosmological horizons provided $m_{\mathrm{ext}}<m<m_{\mathrm{crit}%
}$ an extreme black hole with cosmological horizon if $m_{\mathrm{ext}}=m<m_{%
\mathrm{crit}}$ and a naked singularity with cosmological horizon for the
cases $m<m_{\mathrm{ext}}$ and $m\geq m_{\mathrm{crit}}$\ as one may see in
Fig. \ref{fig10}. As it is clear from Figs. \ref{fig9} and \ref{fig10}, we
have cosmological horizon for $\Lambda <0$. This result is interesting since
it shows that we can have cosmological horizon in the presence of a negative
cosmological constant. 
\begin{figure}[t]
\epsfxsize=7cm \centerline{\epsffile{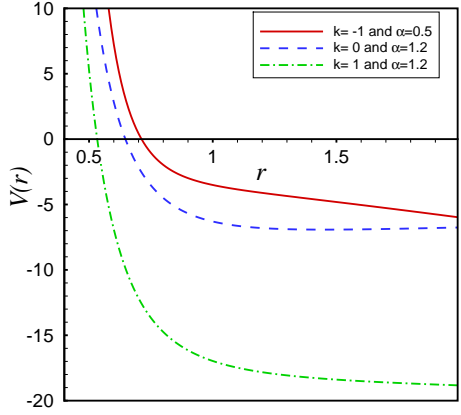}}
\caption{The function $V(r)$ versus $r$ for $n=6$, $p=1.1$, $l=b=1$, $%
\Lambda =15$, $q=1.5$ and $m=0.5$.}
\label{fig8}
\end{figure}
\begin{figure}[t]
\epsfxsize=7cm \centerline{\epsffile{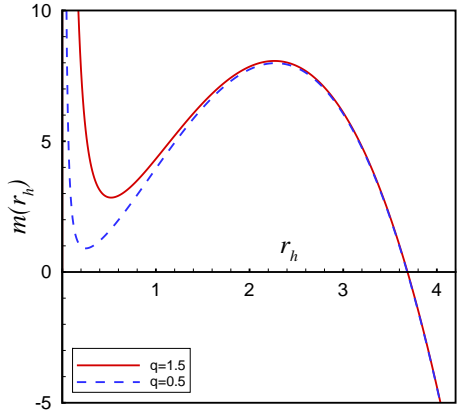}}
\caption{The function $m(r_{h})$ versus $r_{h}$ for $k=1$, $n=6$, $p=1.1$, $%
\protect\alpha =1.18$, $l=1$, $b=1.4$, $\Lambda =-15$.}
\label{fig9}
\end{figure}
\begin{figure}[t]
\epsfxsize=7cm \centerline{\epsffile{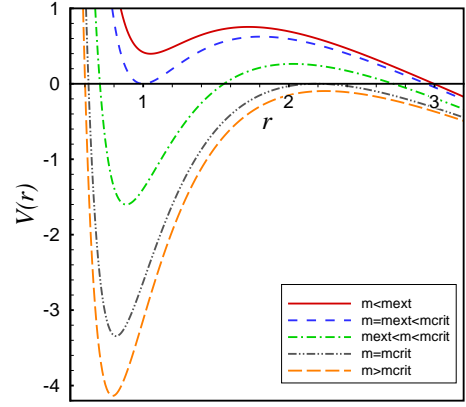}}
\caption{The function $V(r)$ versus $r$ for $k=1$, $n=6$, $p=1.1$, $l=1$, $%
b=1.4$, $\Lambda =-15$, $\protect\alpha =1.18$ and $r_{\mathrm{ext}}=1$.}
\label{fig10}
\end{figure}

\section{THERMODYNAMICS OF TOPOLOGICAL BD BLACK HOLES\label{Ther}}

In this section, we are going to check the validity of the first law of
thermodynamics for the topological BD black holes with power-law Maxwell
field. In order to do this, we should calculate the conserved and
thermodynamic quantities. We start with temperature. The Hawking temperature
of the topological black holes on the outer horizon $r_{+}$ can be
calculated using the relation 
\begin{equation}
T_{+}=\frac{\kappa }{2\pi }=\frac{U^{\prime }(r_{+})}{4\pi \sqrt{U/V}},
\end{equation}%
where $\kappa $ is the surface gravity. One obtains

\begin{eqnarray}
T_{+} &=&\frac{(1+\alpha ^{2})}{4\pi }\left\{ \frac{k\left( n-2\right) }{%
b^{2\gamma }(1-\alpha ^{2})r_{+}^{1-2\gamma }}-\frac{\Lambda b^{2\gamma
}r_{+}^{1-2\gamma }}{n-1}\right.  \notag \\
&&\left. -\frac{2^{p}p\left( 2p-1\right) {b}^{\,-{2\left( n-2\right) \gamma
p/}\left( 2\,p-1\right) }{q}^{2\,p}}{\Pi {r}_{+}^{(2p(n-2)(1-\gamma
)+1)/(2p-1)}}\right\} .  \label{temp}
\end{eqnarray}%
Comparing the temperature (\ref{temp}) with one calculated in the case of
EDPM gravity \cite{KSD}, one finds that temperature is invariant under the
conformal transformation (\ref{Conf}). This comes from the fact that the
conformal parameter at the horizon is regular.

The charge of the black hole can be calculated through the modified Gauss
law, 
\begin{equation}
Q=\frac{\,{1}}{4\pi }\int (rH)^{n-1}(-F)^{p-1}F_{\mu \nu }n^{\mu }u^{\nu }d{%
\Omega }_{n-1},
\end{equation}%
where $n^{\mu }$\ and $u^{\nu }$\ are the unit spacelike and timelike
normals to a sphere of radius $r$\ given as 
\begin{equation*}
n^{\mu }=\frac{1}{\sqrt{-g_{tt}}}dt=\frac{1}{\sqrt{U(r)}}dt,\text{ \ \ \ \ }%
u^{\nu }=\frac{1}{\sqrt{g_{rr}}}dr=\sqrt{V(r)}dr.
\end{equation*}%
Therefore the charge can be computed as 
\begin{equation}
Q=\frac{\,{2^{p-1}{q}^{2\,p-1}}\omega _{n-1}}{4\pi },
\end{equation}%
where $\omega _{n-1}$\ denotes the volume of constant curvature hypersurface 
$h_{ij}dx^{i}dx^{j}$. In addition to temperature and charge we need to
calculate mass, entropy and electric potential in order to check the
satisfaction of the first law of thermodynamics. We can obtain the
Arnowitt-Deser-Misner (ADM) mass $M$, entropy $S$\ and electric potential $U$%
\ of the topological black holes by using the Euclidean action method \cite%
{CaiSu}. In this method, one should first fix the electric potential and the
temperature on the boundary with a fixed radius $r_{+}$. Then, in order to
make the metric positive definite, $t$ should be substituted by $i\tau $: 
\begin{equation}
ds^{2}=U(r)d\tau ^{2}+\frac{1}{V(r)}dr^{2}+r^{2}H^{2}(r)h_{ij}dx^{i}dx^{j}.
\label{Eumetr}
\end{equation}%
This is a necessary step to make the Euclidean action. Since one encounters
a conical singularity at the horizon $r=r_{+}$\ in the Euclidean metric \cite%
{CaiSu}, the Euclidean time $\tau $\ is made periodic with period $\beta $,
where $\beta $\ is the inverse of Hawking temperature in order to eliminate
this singularity. Now, we are ready to obtain the Euclidean action of $(n+1)$%
-dimensional BDPM black hole. The Euclidean action has two parts namely bulk
and surface. The Euclidean action can be calculated analytically and
continuously changing of the action (\ref{Act}) to the Euclidean time $\tau $%
, i.e.,%
\begin{eqnarray}
I_{BDE} &=&-\frac{1}{16\pi }\int_{\mathcal{M}}d^{n+1}x\sqrt{g}\left( \Phi {R}%
-\frac{\omega }{\Phi }(\nabla \Phi )^{2}-V(\Phi )\right.  \notag \\
&&\left. +(-F)^{p}\right) -\frac{1}{8\pi }\int {d^{n}x\sqrt{h}\Phi (K-K_{0})}%
,  \label{act1e}
\end{eqnarray}%
where $K_{0}$\ is the trace of the extrinsic curvature for the boundary
metric $h$ when $q=0$\ and $m=0$. $K_{0}$ is added in order to normalize the
Euclidean action to zero in background \cite{Brown}. Using the metric (\ref%
{Eumetr}), one can obtain%
\begin{eqnarray}
R &=&-g^{-1/2}(g^{1/2}U^{\prime }V/U)^{\prime }-\left( n-1\right) \left[ {%
2nVH^{\prime }/}\left( rH\right) \right.  \notag \\
&&\left. +{2VH^{\prime \prime }/H}+\left( {V^{\prime }/r}+{V^{\prime
}H^{\prime }/H}\right) \right.  \notag \\
&&\left. +\left( n-2\right) \left( {V/r}^{2}+{H^{\prime 2}V/}H^{2}-{k/}%
\left( rH\right) ^{2}\right) \right] ,  \label{RE} \\
K &=&-\frac{\sqrt{V}\left[ rHU^{\prime }+2(n-1)\left( UH+rUH^{\prime
}\right) \right] }{2rHU},  \label{K}
\end{eqnarray}%
Inserting $U(r)$\ and $V(r)$\ from (\ref{U(r)}) and (\ref{V(r)}) with $q=0$\
and $m=0$\ in $K$, one obtains 
\begin{widetext}
\begin{eqnarray}
K_{0} &=&\left( \frac{b}{r}\right) ^{\Gamma /2}\left( \frac{k\left(
n-2\right) \left( {\alpha }^{2}+1\right) ^{2}{b}^{-2\gamma }{r}^{2\gamma }}{(%
{\alpha }^{2}+n-2)({\alpha }^{2}-1)}-\frac{2\,\Lambda \left( {\alpha }%
^{2}+1\right) ^{2}{b}^{{2\gamma }}\,{r}^{2-2\gamma }}{(n-1)({\alpha }^{2}-n)}%
\right) ^{1/2}\,  \notag \\
&&\left\{ 2{b}^{2\gamma }\left( \alpha ^{2}-1\right) \left( n\gamma
-n+4p-1+(-6p+1)\gamma \right) \left( {\alpha }^{2}+n-2\right) n\Lambda {r}%
^{2\left( \,1-\gamma \right) }\right.   \notag \\
&&+\left. \left( n-2\right) \left( n-{\alpha }^{2}\right) \left( n-1\right)
k[\left( n\gamma -n-6p\gamma +4p-1\right) (n-1)-2\left( p-1\right) {\alpha }%
^{2}]{r}^{2\gamma }{b}^{-2\gamma }\right\}   \notag \\
&&{r}^{-1}\left( n-4p+1\right) ^{-1}[2{b}^{2\gamma }\Lambda \,\left( \alpha
^{2}-1\right) \left( {\alpha }^{2}+n-2\right) \,{r}^{2(1-\gamma )}+{r}%
^{2\gamma }{b}^{-2\gamma }k\left( n-1\right) \left( n-2\right) \left( n-{%
\alpha }^{2}\right) ]^{-1}  \label{K0}
\end{eqnarray}%
\end{widetext}Substituting $R$ and $K_{0}$ give by Eqs. (\ref{RE}) and (\ref%
{K0}) in the Euclidean action (\ref{act1e}) and using Eqs. (\ref{U(r)})-(\ref%
{H(r)}), after a long calculation, one can compute the Euclidean action in
terms of the model parameters as 
\begin{eqnarray}
I_{BDE} &=&\frac{\omega _{n-1}}{4}\left\{ \beta \frac{b^{(n-1)\gamma }(n-1)m%
}{4\pi (1+\alpha ^{2})}-\left( b^{(n-1)\gamma }r_{+}^{(n-1)(1-\gamma
)}\right) \right.  \notag \\
&&\left. -\beta \frac{{2^{p}}\left( n-1\right) {p}^{2}{{q}^{2\,p}}}{2\pi {b}%
^{-{\left( 2\,p-n+1\right) \gamma /}\left( 2\,p-1\right) }\Pi \Upsilon {r}%
_{+}^{\Upsilon }}\right\} \,.  \label{IE}
\end{eqnarray}%
On the other hand, we know that the thermodynamic potential can be given by $%
I_{BDE}$ \cite{Brown,Brown2,Brown3,Brown4} 
\begin{equation}
I_{BDE}=\beta M-S-\beta UQ,  \label{GD}
\end{equation}%
where $M$\ is the ADM mass, $S$\ is entropy and $U$\ is electric potential.
One can easily compare Eq. (\ref{IE}) with Eq. (\ref{GD}) and find that%
\begin{equation}
{M}=\frac{b^{(n-1)\gamma }(n-1)m\omega _{n-1}}{16\pi (\alpha ^{2}+1)},
\label{Mass}
\end{equation}%
\begin{equation}
{S}=\frac{b^{(n-1)\gamma }r_{+}^{(n-1)(1-\gamma )}\omega _{n-1}}{4},
\label{entropy}
\end{equation}%
\begin{equation}
U=\frac{\left( n-1\right) {p}^{2}q}{{b}^{-{\left( 2\,p-n+1\right) \gamma /}%
\left( 2\,p-1\right) }\Pi \Upsilon {r}_{+}^{\Upsilon }}.  \label{elect}
\end{equation}%
Here, it is worthwhile to give some remarks. First, it is notable that the
quantities obtained in this section, either conserved or thermodynamic ones,
coincide with those calculated in \cite{KSD}. This fact shows that these
quantities are invariant under the conformal transformation (\ref{Conf}).\
It is also worth to note that although entropy is invariant under conformal
transformations, it does not follow the area law \cite{kang} in contrast
with the case of dilaton black holes in the presence of power-law Maxwell
field \cite{KSD}. This is due to the fact that, in the Euclidean action
formalism the entropy comes from the boundary term. Now, we turn back to the
main purpose of this section which is seeking for satisfaction of first law
of thermodynamics for topological BD black holes. It is a matter of
calculations to check that the conserved and thermodynamic quantities
calculated in this section satisfy the first law of black hole thermodynamics%
\begin{equation}
dM=TdS+Ud{Q}.
\end{equation}%
It is also worth mentioning that since the thermodynamic quantities of our
topological solutions in BDPM gravity coincide with the ones in EDPM
gravity, thermal stability discussions are the same. The stability of the
topological black hole under thermal perturbations in EDPM gravity has been
discussed extensively in \cite{KSD}.

\section{CLOSING REMARKS}

In this paper, we constructed a new class of topological black hole
solutions of BD theory in the presence of a power-law Maxwell field with the
Lagrangian. For this purpose, we introduced the conformal transformation (%
\ref{Conf}) that transforms the EDPM Lagrangian to the BDPM one. Then, by
using this conformal transformation, we obtained BD solutions from
Einstein-dilaton solutions presented in \cite{KSD}. This fact that the BD
scalar field is a localized function accompanied with the fact that the
effects of mass and charge should disappear at infinity restrict the allowed
ranges of the power of electromagnetic source Lagrangian, $p$, and the BD
coupling constant, $\omega $. Consequently, we showed that the permitted
ranges of $p$ and $\omega $ are $1/2<p<(n+1)/4$ and $\omega
>-n/(n-1)+(n-4p+1)(n+1)/4p^{2}(n-1)(n-2)$. In these allowed ranges, our BD
solutions are always well defined, except for $\omega =\left(
(n-4p+1)^{2}-4np^{2}\right) /4p^{2}(n-1)$ in the cases of spherical ($k=1$)
and hyperbolic ($k=-1$) horizons. Also, self-interacting potential $V(\Phi )$
is bounded from below for suitable choices of the model parameters which
guarantees the stability of the system. It is worth mentioning that in the
above ranges of $p$ and $\omega $, the charge term is always dominant in the
vicinity of $r=0$ and positive in metric functions and, therefore, there are
no Schwarzschild-like solutions. Next, we studied the conditions under which
we have black holes with and without cosmological horizon. Interestingly
enough, we showed that our solutions have cosmological horizon even in the
presence of negative cosmological constant.

In order to check the satisfaction of the first law of black holes
thermodynamics, we first computed temperature and charge. Then, by
calculating the Euclidean action, we obtained the black hole's mass, entropy
and electromagnetic potential energy. Using these thermodynamic and
conserved quantities, we showed that the first law of thermodynamics is
satisfied. We found that the entropy does not obey the so-called area law
and the conserved and thermodynamic quantities are invariant under conformal
transformation (\ref{Conf}). All our results recover the results of \cite%
{Confsol3} for linear Maxwell field in the limiting case where $p=1$.

Finally, we would like to mention that in this paper we obtained static
topological black hole solutions of BD gravity in the presence of power-law
Maxwell nonlinear electrodynamics. For future studies, one can extend the
studies to the rotating black holes/branes in BD gravity with power-law
Maxwell source. One may also consider other nonlinear electromagnetic
sources such as Born-Infeld, logarithmic and exponential Lagrangian and
obtain the black hole solutions of these theories in the framework of BD
gravity.

\acknowledgments{We thank Shiraz University Research Council. This
work has been supported financially by the Research Institute for
Astronomy and Astrophysics of Maragha, Iran.}

\end{document}